\documentstyle[12pt]{article}

\def\sdspace{\baselineskip = .25in}

\def\Tilde#1{\widetilde {#1}}

\begin{document}
\title{Predictive Neutrino Spectrum in Minimal SO(10) Grand
Unification}

\author{{\bf K.S. Babu\thanks{Supported in part by Department of Energy Grant
\#DE-FG02-91ER406267}}\\
Bartol Research Institute\\
University of Delaware\\ Newark, DE 19716 \\ and \\
{\bf R.N. Mohapatra\thanks{Supported by National
Science Foundation Grant PHY-9119745}}\\
Department of Physics\\
University of Maryland\\ College Park, MD 20742}

\date {}
\maketitle
\begin{abstract}

We show that minimal SO(10) Grand Unification models where the fermions
have Yukawa couplings to only one (complex) {\bf 10} and
one {\bf 126} of Higgs scalars lead to a very predictive neutrino spectrum.
This comes about since the standard model doublet contained in the
{\bf 126} of Higgs (needed for the see--saw mechanism)
receives an induced vacuum expectation value at tree--level, which,
in addition to correcting the bad
asymptotic mass relations $m_d=m_e$ and $m_s=m_\mu$,
also relates the Majorana neutrino mass matrix to observables in the
charged fermion sectors.  We
find that (i) the $\nu_e-\nu_{\mu}$ mixing angle relevant for the solar
neutrinos can be considerably smaller than
the Cabibbo angle and lies in the range
${\rm sin}\theta_{e \mu}= 0-0.3$, (ii) $\nu_e-\nu_\tau$ mixing is
sin$\theta_{e \tau} \simeq 3|V_{td}| \simeq 0.05$,
(iii) the $\nu_\mu-\nu_\tau$ mixing angle is large,
${\rm sin}\theta_{\mu \tau} \simeq 3|V_{cb}|=0.12-0.16$, and (iv)
$m_{\nu_\tau}/m_{\nu_\mu} \ge 10^3$, implying that $\nu_{\mu}-\nu_\tau$
oscillations should be accessible to forthcoming experiments.

\end{abstract}
\newpage
\sdspace

It is quite possible that the deficit of solar neutrinos reported in the
Chlorine,$^1$ Kamiokande,$^2$ SAGE,$^3$ and GALLEX$^4$
experiments is an indication that
the neutrinos have masses and mixings very much like the quarks.
The observed deficit can be explained in terms of neutrino oscillations in two
different ways: (i) long wave length vacuum oscillation,$^5$ and (ii)
resonant matter oscillation (the Mikheyev-
Smirnov-Wolfenstein (MSW) effect$^6$).
Assuming a two-flavor $\nu_e-\nu_\mu$ oscillation,
in the former case, the neutrino masses
and mixing angle should satisfy$^5$ (at 90\% CL)
$\Delta m^2  \simeq (0.5~
{\rm to}~1.1)\times
10^{-10}~eV^2$ and sin$^22\theta_{e\mu} \simeq (0.75~ {\rm to}~1)$.  In case
of MSW, on the other hand, there are two allowed windows that fit
all of the experimental data$^7$:
(a) the small mixing angle non--adiabatic
solution, which requires $\Delta m^2 \simeq (0.3~ {\rm to}~1.2)\times
10^{-5}~eV^2$ and ${\rm sin}^22\theta_{e\mu} \simeq (0.4~{\rm to}~1.5)
\times 10^{-2}$, and
(b) the large angle solution with $\Delta m^2\simeq (0.3~{\rm to}~5)
\times 10^{-5}~eV^2$
and ${\rm sin}^22\theta_{e\mu} \simeq (0.5~{\rm to}~0.9)$.
In all these cases, barring an unlikely scenario of near mass degeneracy among
neutrinos, either $\nu_\mu$ or $\nu_\tau$ should have mass in the
$(10^{-5}~{\rm to} ~10^{-3})~eV$ range.

A well--known and elegant explanation for the origin of such tiny neutrino
masses is the see--saw mechanism,$^8$ wherein the light neutrino
masses scale inversely with the $B-L$ breaking Majorana mass $M$ of the
right--handed neutrino:
$m_\nu \simeq m^2/M$, $m$ being the neutrino Dirac mass.
The smallness of $m_\nu$ is then understood in
terms of the heaviness of the $B-L$ breaking scale.  The solar neutrino
puzzle indicates that the $B-L$ scale is in the
$(10^{12}-10^{16})~GeV$ range.

All of the observations above, viz., non--zero neutrino masses, the see--saw
mechanism, and a high $B-L$ scale, fit rather naturally in grand
unified models based on the gauge group $SO(10)^{9}$.
In its non-supersymmetric version, experimental
constraints from proton life--time and the weak mixing angle sin$^2\theta_W$
require that $SO(10)$ breaks not directly into the standard model, but at
least in two steps.  In a two--step breaking scheme, the left--right
symmetric
intermediate scale is around $10^{12}$ GeV.$^{10}$
In supersymmetric $SO(10)$, on the other hand, there is no need for an
intermediate scale, $SO(10)$ can break directly to the
standard model at around $10^{16}$ GeV.

To confront $SO(10)$ models with the solar neutrino data, one must make
precise predictions of the neutrino masses and mixing angles.  This
requires, however, detailed
information of the Dirac neutrino mass matrix as well as the Majorana
matrix.  In grand unified theories (Guts), it is possible to
relate the
quark masses with the lepton masses.  For example, in
$SU(5)$ Gut, the charge $-1/3$ quark mass matrix is
equal to the charged
lepton matrix at the unification scale.  Most $SO(10)$ models retain
this feature; in addition, they also relate
the neutrino Dirac
mass matrix
to the charge $2/3$ quark matrix.   However, there is no
simple way, in general, to relate the heavy Majorana matrix $M$ to the
charged fermion observables.

The purpose of this Letter is to show that in a class of minimal $SO(10)$
models, in fact, not only the Dirac neutrino matrix, but the Majorana
matrix also gets related to observables in the charged fermion sector.
This leads to a very predictive neutrino spectrum, which we analyze.
We use a simple Higgs system with
one (complex) {\bf 10}
and one {\bf 126} that have Yukawa couplings to fermions.
The {\bf 10} is needed for
quark and lepton masses, the
{\bf 126} is needed for the see--saw mechanism.  Crucial to the
predictivity of the neutrino spectrum is the observation that
the standard model doublet contained in the {\bf 126}
receives an induced vacuum expectation
value (vev) at tree--level.  In its absence, one would
have the asymptotic mass
relations $m_b=m_\tau,~m_s=m_\mu,~m_d=m_e$, as in minimal $SU(5)$.
While the first relation would lead to
a successful prediction of $m_b$ at low energies, the last two
are in disagreement with observations.  The induced vev of the standard
doublet of {\bf 126} corrects these bad relations and at the
same time
also relates the Majorana neutrino mass matrix to
observables in the charged fermion sector, leading to a predictive
neutrino spectrum.

We shall consider non--Susy $SO(10)$ breaking to the standard model via
the $SU(2)_L \times SU(2)_R
\times SU(4)_C \equiv G_{224}$ chain as well as
Susy-$SO(10)$ breaking directly to the standard model.
Our predictions on the neutrino mass ratios and the mixing
angles are essentially
independent of the chain of descent, it only affects the
overall scale of neutrino masses.

The breaking of $SO(10)$ via $G_{224}$ is achieved by either
a {\bf 54} or
a {\bf 210} of Higgs.  The {\bf 210} also breaks
the discrete $D$--parity,$^{11}$ the {\bf 54} preserves it.
$D$--parity is a local discrete
$Z_2$--subgroup of $SO(10)$, under $D$, a fermion field $f$ transforms
into its charge conjugate $f^c$.  Breaking of $D$--parity at the Gut
scale makes the see--saw mechanism natural.$^{12}$
The second stage of symmetry breaking
goes via the {\bf 126}.  Finally, the electro--weak symmetry breaking
proceeds via the {\bf 10}.
Note that in Susy-$SO(10)$, the
first two symmetry breaking scales coalesce into one.

Let us turn attention to the fermion--Higgs Yukawa couplings of the
model.  Denoting
the three families of fermions belonging to
{\bf 16}--dimensional spinor representation of $SO(10)$ by
$\psi_a$, $a=1-3$, the complex {\bf 10}--plet of Higgs by $H$, and
the {\bf 126}--plet of Higgs by $\Delta$,
the Yukawa couplings can be written down as
\begin{equation}
L_Y = h_{ab}\psi_a\psi_bH + f_{ab}\psi_a\psi_b\overline{\Delta} + H.C.
\end{equation}
Note that since the {\bf 10}--plet is complex, one other coupling
$\psi_a\psi_b\overline{H}$ is allowed in general.  In Susy--$SO(10)$, the
requirement of supersymmetry prevents such a term.  In the non--Susy
case, we forbid this term by imposing a $U(1)_{PQ}$
symmetry, which may anyway be needed in order to solve the strong CP
problem.

The {\bf 10} and {\bf 126} of Higgs have the following decomposition
under $G_{224}$:
\begin{eqnarray}
{\bf 126} &\rightarrow& (1,1,6)+(1,3,10)+(3,1,\overline{10})+(2,2,15)
\nonumber\\
{\bf 10} &\rightarrow& (1,1,6)+(2,2,1)
\end{eqnarray}
Denote the $(1,3,10)$ and $(2,2,15)$ components of
$\Delta({\bf 126})$
by $\Delta_R$ and
$\Sigma$ respectively and the $(2,2,1)$ component of $H({\bf 10})$ by
$\Phi$.
The vev $<\Delta_R^0> \equiv v_R \sim 10^{12}~GeV$
breaks the intermediate symmetry down to
the standard model and generates Majorana
neutrino masses given by $fv_R$.
$\Phi$  contains two standard model doublets
which acquire
non--zero vev's denoted by $\kappa_u$ and $\kappa_d$ with
$\kappa_{u,d} \sim 10^{2}~GeV$.
$\kappa_u$ generates charge 2/3 quark as well as Dirac neutrino
masses, while $\kappa_d$ gives rise to $-1/3$ quark and charged lepton
masses.

Within this minimal picture, if $\kappa_u,~\kappa_d$ and $v_R$ are
the only vev's
contributing to fermion masses, in addition to
the $SU(5)$ relations $m_b=m_\tau,~
m_s=m_\mu,~m_d=m_e$, eq. (1) will also lead to the unacceptable relations
$m_u:m_c:m_t = m_d:m_s:m_b$.
Moreover, the
Cabibbo-Kobayashi-Maskawa (CKM) mixing matrix will be identity.
Fortunately, within this minimal scheme, we have found that
there are new contributions
to the fermion mass matrices which are just of the right order of
magnitude to correct these bad relations.  To see this, note that
the scalar potential contains, among other terms, a crucial term
\begin{equation}
V_1 =\lambda \Delta \overline{\Delta} \Delta H +H.C.
\end{equation}
Such a term is invariant under the $U(1)_{PQ}$ symmetry.  It will be
present in the Susy $SO(10)$ as well, arising from the {\bf 210}
$F$--term.
This term induces vev's for the standard doublets contained in the $\Sigma$
multiplet of {\bf 126}.  The vev arises through a term
$\overline{\Delta}_R\Delta_R \Sigma \Phi$ contained in $V_1$.
This induced vev can also be seen by
analyzing the one-loop graph involving neutrinos
which generates a divergent contribution to such a term.

We can estimate the magnitudes of the induced vev's
of $\Sigma$ (denoted by $v_u$ and $v_d$ along the up
and down directions) assuming the survival hypothesis to hold:
\begin{equation}
v_{u,d} \sim \lambda \left({v_R^2}\over {M_{\Sigma}^2}\right)
\kappa_{u,d}~~.
\end{equation}
Suppose $M_U \sim 10^{15}~GeV$, $M_I \sim 3 \times 10^{12}~GeV$
and $M_{\Sigma} \sim 10^{14}~
GeV$, consistent with survival hypothesis, then $v_u$ and
$v_d$ are of order 100 MeV, in the right range for correcting the bad
mass relations.  We emphasize that there is no need for a
second fine--tuning to generate such induced vev's.  In the Susy
version, since there is no intermediate scale at all, the
factor $(v_R^2/M_{\Sigma}^2)$ is not a suppression, so
the induced vev's
can be as large as $\kappa_{u,d}$.

We are now in a position to write down the quark and lepton mass
matrices of the model:
\begin{eqnarray}
M_u = h \kappa_u +fv_u~~&~&~~M_d = h \kappa_d+f v_d \nonumber \\
M_{\nu}^D = h \kappa_u-3 f v_u~~&~&~~M_l=h \kappa_d-3fv_d\nonumber \\
M_{\nu}^M &=& f v_R~.
\end{eqnarray}
Here $M_{\nu}^D$ is the Dirac neutrino matrix and $M_{\nu}^M$ is the
Majorana mass matrix.

Before proceeding, we should specify the origin of CP violation in the
model.
We shall assume that it is spontaneous or soft, that will keep the
number of parameters at a minimum.
The Higgs sector described above already has enough structure to
generate realistic CP violation either softly or spontaneously.$^{13}$
The Yukawa
coupling matrices $h$ and $f$ in this case
are real and symmetric.  Although there
will be three different phases in the vev's (one common phase for
$\kappa_u$ and $\kappa_d$ and one each for $v_u$ and $v_d$),
only two combinations enter into the mass matrices,
as the overall phase can be removed from
each sector.  We shall bring these two phases into $v_u$ and $v_d$
and hence
forth denote them by $v_ue^{i \alpha}$ and $v_d e^{i \beta}$.

To see the predictive power of the model as regards the neutrino
spectrum, note that we can choose a basis
where one
of the coupling matrices, say $h$, is real and diagonal.  Then there are
13 parameters in all, not counting the superheavy scale $v_R$: 3
diagonal elements of the matrix $h \kappa_u$, 6 elements of $f v_u$, 2
ratios of vev's $r_1=\kappa_d/\kappa_u$ and $r_2=v_d/v_u$, and the two
phases $\alpha$ and $\beta$.
These 13 parameters are related to the 13 observables in the
charged fermion sector, viz., 9 fermion masses, 3 quark mixing
angles and one CP violating phase.  The light neutrino mass matrix will
then be completely specified in terms of other physical observables
and the overall scale $v_R$.  That would lead to 8
predictions in the lepton sector:  3 leptonic
mixing angles, 2 neutrino mass ratios and 3 leptonic CP violating
phases.

The relations of eq. (5) hold at the intermediate
scale $M_I$ where quark--lepton
symmetry and left--right symmetry are intact.  There are calculable
renormalization corrections to these relations below $M_I$.  The quark and
charged lepton masses as well as the CKM matrix elements run between $M_I$ and
low energies.  The neutrino masses and mixing angles, however, do not
run below $M_I$, since the right-handed neutrinos have masses of order
$M_I$ and decouple below that scale.  The predictions in the neutrino
sector should then be arrived at by first extrapolating the charged fermion
observables to $M_I$.

We shall present results for the non--Susy $SO(10)$ model with the
$G_{224}$ intermediate symmetry.  We fix
the intermediate scale at $M_I = 10^{12}~GeV$ and use the one--loop
standard model renormalization group equations to track the running of
the gauge couplings between $M_Z$ and $M_I$.
For Susy--$SO(10)$, the results are
similar, we shall postpone details to a forthcoming longer paper.$^{14}$

To compute the renormalization factors, we choose as low energy inputs
the gauge couplings at $M_Z$ to be
\begin{eqnarray}
\alpha_1 (M_Z) = 0.01688~~;~~\alpha_2 (M_Z) = 0.03322~~;~~\alpha_3 (M_Z)
=0.12~~.
\end{eqnarray}
For the light quark (running) masses, we choose values listed in Ref.
(15):
\begin{eqnarray}
m_u (1~GeV)=5.1\pm 1.5 ~MeV~ &~& ~~m_d (1~GeV)=8.9 \pm 2.6 MeV \nonumber
\\
m_s (1~GeV)=175 \pm 55~MeV ~ &~& ~~
m_c (m_c) = 1.27 \pm 0.05~GeV \nonumber \\
{}~m_b (m_b) &=& 4.25 \pm 0.1~GeV~~.
\end{eqnarray}
The top--quark mass will be allowed to vary between 100 and 200 GeV.

Between 1 GeV and $M_Z$, we use two--loop QCD renormalization group
equations for the running of the quark masses and the $SU(3)_C$
gauge coupling,$^{15}$ treating
particle thresholds as step functions.  From $M_Z$ to $M_I$, the
running factors are computed semi--analytically both for the fermion
masses and for the CKM angles
by using the
one--loop renormalization group equations for the Yukawa couplings
and keeping the heavy top--quark contribution.$^{16}$  The running factors,
defined as $\eta_i = m_i(M_I)/m_i(m_i)$
[$\eta_i=m_i(M_I)/m_i(1~GeV)$
for light quarks $(u,d,s)$] are $\eta (u,c,t)=(0.227,0.241, 0.452)$,
$\eta (d,s,b)=$ $(0.232,0.232,0.295)$,
$\eta (e,\mu,\tau)=0.912$ for the case of
$m_t=130~GeV$ and $\eta (u,c,t)=(0.237,0.252,0.482)$,
$\eta (d,s,b)=(0.242,0.242,0.302),~\eta (e,\mu,\tau)=0.952$ for $m_t=150~GeV$.
The (common) running factors for
the CKM angles (we follow the parameterization advocated by the Particle
Data Group)
$S_{23}$ and $S_{13}$ are 1.054 for $m_t=130~GeV$ and
$1.077$ for $m_t=150~GeV$.  The Cabibbo angle $S_{12}$ and the KM phase
$\delta_{KM}$ are essentially unaltered.

Let us first analyze the mass matrices of eq. (5) in the limit of CP
conservation.  We shall treat spontaneous CP violation arising through
the phases of the vev's $v_ue^{i \alpha}$ and $v_de^{i \beta}$ as small
perturbations.  This procedure will be justified a posteriori.  In fact,
we find that realistic fermion masses, in particular the first family
masses, require these phases to be small.

We can rewrite the mass matrices $M_l, M_\nu^D$ and $M_{\nu}^M$ of eq.
(5) in terms of the quark mass matrices and three ratios of
vev's -- $r_1=\kappa_d/\kappa_u,~r_2=v_d/v_u~,R=v_u/v_R$:
\begin{eqnarray}
M_l &=& {{4 r_1 r_2}\over{r_2-r_1}} M_u-{{r_1+3 r_2}\over
{r_2-r_1}} M_d ~~,\nonumber \\
M_\nu^D &=& {{3 r_1+r_2}\over {r_2-r_1}} M_u-{4 \over {r_2-r_1}}M_d ~~,
\nonumber \\
M_\nu^M &=& {1 \over R} {{r_1} \over {r_1-r_2}}
M_u -{1 \over R}{1 \over {r_1-r_2}}M_d~~.
\end{eqnarray}
It is convenient to go to a basis where $M_u$ is diagonal.  In that
basis, $M_d$ is given by $M_d=V M_d^{diagonal} V^T$, where
$M_d^{diagonal}={\rm diagonal}(m_d,m_s,m_b)$ and $V$ is the CKM matrix.  One
sees that $M_l$ of eq. (8) contains only physical observables from the
quark sector and two parameters $r_1$ and $r_2$.  In the CP--conserving
limit then, the three eigen--values of $M_l$ will lead to one mass prediction
for the charged fermions.  To
see this prediction, $M_l$ needs to be diagonalized.  Note first that by
taking the Trace of $M_l$ of eq. (8), one obtains a
relation for $r_1$ in terms of
$r_2$ and the charged fermion masses.  This is approximately
$r_1 \simeq \left(m_\tau+3 m_b\right)/4 m_t$ (as long as
$r_2$ is larger than $m_b/m_t$).
Since $|m_b| \simeq |m_\tau|$ at the
intermediate scale to within 30\% or so, depending on the relative sign
of $m_b$ and $m_\tau$, $r_1$ will be close to either $m_b/m_t$ or
to $(m_b/2m_t)$.
Note also that if $r_2 \gg r_1$, $M_l$ becomes
independent of $r_2$, while $M_{\nu}^D$ retains some dependence:
\begin{equation}
M_l \simeq 4 r_1 M_u -3 M_d~~,~~
M_{\nu}^D \simeq M_u-{4 \over {r_2}}M_d~~.
\end{equation}
This means that the parameter $r_2$ will only be
loosely constrained from the charged fermion sector.

We do the fitting as follows.
For a fixed value of $r_2$, we determine $r_1$ from the Tr$(M_l)$
using the input
values of the masses and the renormalization factors
discussed above.  $M_l$ is then diagonalized
numerically.  There will be two mass relations among charged fermions.
Since the charged lepton masses are precisely known at low energies,
we invert these relations to predict the $d$--quark and $s$--quark
masses.  The $s$--quark mass is sensitive to the muon
mass, the $d$--mass is related to the electron mass.
This procedure is repeated for other values of
$r_2$.  For each choice, the light neutrino masses and the leptonic CKM
matrix elements are then computed using the see--saw formula.

We find that there are essentially three different solutions.  A
two--fold ambiguity arises
from the unknown relative sign of $m_b$ and $m_\tau$ at $M_I$.
We have found acceptable solutions for both signs.
Our numerical fit shows that the loosely constrained parameter
$r_2$ cannot be
smaller than 0.1 or so, otherwise the $d$--quark mass comes out
too small.  Now, the light neutrino spectrum is sensitive to
$r_2$ only when $r_2 \sim 4 m_s/m_c\sim \pm 0.4$, since the two terms in
$M_\nu^D$ of eq. (9)
become comparable (for the second family) then.  For larger values of
$r_2$, the first term in $M_\nu^D$ dominates and the light neutrino
spectrum becomes independent of $r_2$.  Two qualitatively
different solutions are
obtained depending on whether $r_2$ is near $\pm 0.4$ or not.

Numerical results for the three different cases are presented below.
The input values of the CKM mixing angles are chosen for all cases to be
$S_{12}=-0.22,~S_{23}=0.052,~S_{13}=6.24 \times 10^{-3}$.
Since $\delta_{KM}$ has been set to zero for now, we have allowed for
the mixing
angles to have either sign.  Not all signs result in acceptable
quark masses though.  Similarly, the fermion masses can have either sign,
but these are also restricted.
The most stringent constraint comes from the $d$--quark
mass, which has a tendency to come out too small.  Acceptable solutions are
obtained when $\theta_{23},~\theta_{13}$ are
in the first quadrant and $\theta_{12}$ in the fourth quadrant.
We use the precisely known charged lepton masses at low energies and the
running factors discussed above
to arrive at the values of the masses and mixing
angles at $M_I$.

Solution 1:
\begin{eqnarray}
{\rm Input}:  m_u(1~GeV) &=& 4~MeV,~~m_c(m_c)=1.22~GeV,~~m_t =
150~GeV \nonumber \\
m_b(m_b)&=&-4.35~GeV,~~r_1=-1/51.7,~~r_2=2.0 \nonumber \\
{\rm Output}:  m_d(1~GeV) &=& 6.4~ MeV,~~m_s(1 GeV)=164~MeV \nonumber \\
\left(m_{\nu_e},~m_{\nu_{\mu}},~m_{\nu_\tau}\right) &=& R\left(2.0 \times
10^{-2},8.7,-2.2 \times 10^4\right)~GeV \nonumber \\
V_{KM}^{\rm lepton} &=& \left(\matrix{0.9522 & 0.3051 & 0.0123 \cr
-0.2991 & 0.9400 & -0.1637 \cr -0.0615 & 0.1522 & 0.9864}\right)~~.
\end{eqnarray}

Solution 2:
\begin{eqnarray}
{\rm Input}: m_u(1~GeV)&=& 4~MeV,~~m_c(m_c)=1.22~GeV,~~m_t=150~GeV\nonumber
\\
m_b(m_b) &=& -4.35~GeV,~~r_1=-1/51.4,~~r_2=0.2 \nonumber \\
{\rm Output}: m_d(1~GeV) &=& 5.6~MeV,~~ m_s(1~GeV)=175~MeV \nonumber \\
\left(m_{\nu_e},m_{\nu_\mu},m_{\nu_\tau}\right) &=& R\left(6.3 \times
10^{-4},2.1,-2.7 \times 10^3\right)~GeV \nonumber \\
V_{KM}^{\rm lepton} &=& \left(\matrix{0.9969 & 0.0506 & -0.0607 \cr
-0.0585 & 0.9890 & -0.1359 \cr 0.0532 & 0.1390 & 0.9889}\right)~~.
\end{eqnarray}

Solution 3:
\begin{eqnarray}
{\rm Input:} m_u(1~GeV) &=& 3.5~MeV,~m_c(m_c) = 1.27~GeV,~m_t=130~GeV
\nonumber \\
m_b(m_b) &=& -4.35~GeV,~~r_1=-1/101.8,~~r_2=-0.5 \nonumber \\
{\rm Output}: m_d(1~GeV) &=& -5.24~MeV,~~ m_s(1~GeV)=-173~ MeV \nonumber \\
\left(m_{\nu_e},m_{\nu_\mu},m_{\nu_\tau}\right) &=& R\left(4.1 \times
10^{-2},1.1,6.2 \times 10^3\right)~GeV \nonumber \\
V_{KM}^{\rm lepton} &=& \left(\matrix{0.9954 & -0.0794 & -0.0521 \cr
0.0704 & 0.9853 & -0.1559 \cr 0.0637 & 0.1515 & 0.9864}\right)~~.
\end{eqnarray}

Solution 1 corresponds to choosing $r_1 \sim m_b/m_t$.  All the charged
lepton masses are negative in this case.  Since $r_2$ is large, the Dirac
neutrino matrix is essentially $M_u$, which is diagonal; so is the
Majorana matrix.  All the
leptonic mixing angles arise from the charged lepton sector.  Note that
the predictions for $m_d$ and $m_s$ are within the range quoted
in eq. (7).  The mixing angle sin$\theta_{\nu_e-\nu_\mu}$ relevant for solar
neutrinos is 0.30, close to the Cabibbo angle.  Such a value may already
be excluded by a combination of all solar neutrino data taken at the 90\% CL
(but not at the 95\% CL).$^7$  Actually, within the model, there is a
more stringent constraint.  Note that the $\nu_\mu-\nu_\tau$ mixing
angle is large, it is approximately $3|V_{cb}| \simeq 0.16$.  For that
large a mixing, constraints from $\nu_\mu-\nu_\tau$
oscillation experiments imply$^{17}$ that $|m_{\nu_\tau}^2-
m_{\nu_{\mu}}^2| \le 4~eV^2$.  Solution 1 also has
$m_{\nu_\tau}/m_{\nu_\mu} \simeq 2.5 \times 10^3$, requiring that
$m_{\nu_{\mu}} \le 0.8 \times 10^{-3}~eV$.  This is a factor of 2 too
small for $\nu_e-\nu_\mu$ MSW oscillation for the
solar puzzle (at the 90\% CL), but perhaps is not excluded completely,
once astrophysical uncertainties are folded in.
If $\nu_\tau$ mass is around $2 \times 10^{-3}~eV$,
$\nu_e-\nu_{\tau}$ oscillation may be relevant, that mixing angle is
$\simeq 3|V_{td}| \simeq 6\%$.  It would
require the parameter $R=v_u/v_R \sim 10^{-16}$ or $v_R \sim 10^{16}~
GeV$ for $v_u \sim 1~GeV$.  Such a scenario fits well within
Susy--$SO(10)$, but not in the non--Susy $G_{224}$
chain.  Note that $\nu_\tau$ mass is negative, a transformation
$\nu_\tau \rightarrow i \nu_\tau$ will make it positive.

Solution 2 differs from 1 in that $r_2$ is smaller,
$r_2=0.2$.  The $1-2$ mixing in the
neutrino sector is large in this case, so it can cancel the
Cabibbo like mixing arising from the charged lepton sector.  As we vary
$r_2$ from around 0.2 to 0.6, this cancellation becomes stronger, the
$\nu_e-\nu_\mu$ mixing angle becoming zero for a critical value of
$r_2$.  For larger $r_2$, the solution will approach Solution 1.
The
$\nu_\mu-\nu_\tau$ mixing angle is still near $3 |V_{cb}|$, so as
before,  $m_{\nu_{\tau}} \le 2~ eV$.  From the $\nu_\tau/\nu_\mu$ mass
ratio, which is $1.3 \times 10^3$ in this case, we see that
$m_{\nu_\mu} \le 1.6 \times 10^{-3}~eV$.  This is just within the
allowed range$^7$ (at 95\% CL) for small angle non--adiabatic
$\nu_e-\nu_\mu$ MSW oscillation, with a predicted count rate of about
50 SNU for the Gallium experiment.
Note that there is a lower limit of about 1 eV for
the $\nu_\tau$ mass in this case.  Forthcoming experiments should then
be able to observe $\nu_\mu-\nu_\tau$ oscillations.  A $\nu_\tau$ mass
in the (1 to 2) eV range can also be cosmologically significant, it can
be at least part of the hot dark matter.
In Susy $SO(10)$, $\nu_e-\nu_\tau$ oscillation (the relevant mixing is
about $3|V_{td}| \simeq 5\%$), could account for the solar neutrino puzzle.

Solution 3 corresponds to choosing $r_1 \sim (m_b/2m_t)$.
All charged lepton masses are positive in this case.
The sign of
$r_2$ has been chosen to get small sin$\theta_{e-\mu}$.  (For other
values of $r_2$, the results are similar to Solution 1.)
However, the mass
ratio $\nu_\tau/\nu_\mu$ is $\sim 6 \times 10^3$, and
sin$\theta_{\mu \tau} \simeq 3 V_{cb}$
so $\nu_e-\nu_\mu$ oscillation
cannot be responsible for solar MSW.  As in other cases,
$\nu_e-\nu_\tau$ MSW oscillation with a 6\% mixing is a viable possibility.

Observe that none of the solution generates $\nu_e-\nu_\mu$ mixing large
enough for the vacuum oscillation for solar neutrinos.
Similarly, the puzzle with atmospheric
neutrinos cannot be explained in this minimal scheme in terms of
$\nu_\mu-\nu_\tau$ oscillation, the relevant mixing is not large
enough.  (For an $SO(10)$--based explanation of this phenomenon, see
Ref. 18).

Let us now re-instate the CP--violating
phases $\alpha$ and $\beta$ in the vev's perturbatively.  Small
values of the phases are sufficient to account for realistic CP violation
in the quark sector.
We shall present details for the
case of Solution 2 only, others are similar.  We also
tried to fit all the charged
fermion masses and mixing angles for large phases, but found no
consistent solution.

First we make a basis transformation to go from the basis where $M_u$
is diagonal to one where the matrix
$h \kappa_u$ is diagonal.  It is easier to introduce phases in that
basis.
For $\alpha=3.5^0,~\beta=4.5^0$, the
CP--violating parameter $J$ for the quark system$^{19}$ is $J \simeq 1 \times
10^{-5}$, which is sufficient to accommodate $\epsilon$ in the
neutral $K$ system.  The leptonic CP violating phases are
correspondingly small, for eg., the analog of $J$ is
$J_l \simeq 7 \times 10^{-5}$.
These small phases modify the first family masses
slightly, but the effect is less than 10\%.  Our predictions for the neutrino
mixing angles are essentially unaltered.

In summary, we have presented a class of minimal $SO(10)$ grand unified
models where the light neutrino masses and mixing angles are predictable
in terms of observables in the charged fermion sector.  Our approach
here has been orthogonal to some other recent attempts $^{20,18}$ based
on grand unification, we have kept the Higgs sector as simple as
possible and followed its consequences.  We have found three different
types of solutions for the neutrino spectrum.
In Solution 1, the $\nu_e-\nu_\mu$ mixing
angle is near the Cabibbo angle, while Solutions 2 and 3 have it much
smaller.  In all cases, $\nu_e-\nu_\tau$ mixing angle is predicted to be
near $3|V_{td}| \simeq 0.05$ and
$\nu_\mu-\nu_\tau$ mixing angle is
$\simeq 3 |V_{cb}| \simeq 0.15$ with the mass ratio
$m_{\nu_\tau}/m_{\nu_\mu} \ge 10^3$.  If the solar neutrino puzzle is
due to small angle non--adiabatic MSW, as in Solution 2,
$\nu_\mu-\nu_\tau$ oscillation should be observable in the
forthcoming experiments.

\section*{Acknowledgments:}  We wish to thank Kari Enqvist and
the organizers of the 1992 Neutrino Program at Nordita where this work
began for their warm hosptality.  K.S.B wishes to acknowledge Aspen
Center for Physics where part of this work was
done for its hospitality.  He would also like to thank Stuart Raby
and Lincoln Wolfenstein for discussions.
The numerical fits were obtained using Mathematica, R.N.M wishes
to thank Gerry Gilbert and Lee--Dai Gyu for introducing him to
Mathematica.

\section*{References}

\begin{enumerate}
\item R. Davis et. al., in {\it Proceedings of the 21st International
Cosmic Ray Conference}, Vol. 12, ed. R.J. Protheroe (University of
Adelide Press, Adelide, 1990) p. 143.
\item K.S. Hirata et. al., {\it Phys. Rev. Lett.} {\bf 65}, 1297
(1990).
\item A.I. Abazov et. al., {\it Phys. Rev. Lett.} {\bf 67}, 3332 (1991).
\item P. Anselman et. al., {\it Phys. Lett.}
{\bf B285}, 376 (1992).
\item For recent updates of this old idea, see for example:\\
A. Acker, S. Pakvasa and J. Pantaleone, {\it Phys. Rev.} {\bf D 43}, 1754
(1991);\\
P.I. Krastev and S.T. Petcov, {\it Phys. Lett.} {\bf B285}, 85 (1992);\\
V. Barger, R. Phillips and K. Whisnant, University of Wisconsin Preprint
MAD/PH/709 (1992).
\item L. Wolfenstein, {\it Phys. Rev.} {\bf D 17}, 2369 (1978); \\
S.P. Mikheyev and A. Yu Smirnov, {\it Yad. Fiz.} {\bf 42}, 1441 (1985)
[{\it Sov. J. Nucl. Phys}, {\bf 42},
913 (1985)].
\item For recent updates after GALLEX results, see for example:\\
X. Shi, D.N. Schramm and J. Bahcall, {\it Phys. Rev. Lett.} {\bf 69},
717 (1992);\\
S.A. Bludman, N. Hata, D.C. Kennedy and P. Langacker, Pennsylvania
Preprint UPR-0516T (1992).
\item M. Gell-Mann, P. Ramond and R. Slansky, in {\it Supergravity}, ed.
F. van Nieuwenhuizen and D. Freedman (North Holland, 1979), p. 315; \\
T. Yanagida, in {\it Proceedings of the Workshop on Unified Theory and
Baryon Number in the Universe}, ed. A. Sawada and H. Sugawara, (KEK,
Tsukuba, Japan, 1979);\\
R.N. Mohapatra and G. Senjanovic {\it Phys. Rev. Lett.} {\bf 44}, 912
(1980).
\item H. Georgi, in {\it Particles and Fields}, ed. C.E. Carlson, AIP,
New York (1974); \\
H. Fritzsch and P. Minkowski, {\it Ann. Phys.} {\bf 93}, 193 (1975).
\item D. Chang, R.N. Mohapatra, M.K. Parida, J. Gipson and R.E. Marshak,
{\it Phys. Rev} {\bf D 31}, 1718 (1985);\\
For a more recent analysis, see N.G. Deshpande, R. Keith and P.B. Pal,
Oregon Preprint OITS-484 (1992).
\item D. Chang, R.N. Mohapatra and M.K. Parida, {\it Phys. Rev. Lett.}
{\bf 52}, 1072 (1982).
\item D. Chang and R.N. Mohapatra, {\it Phys. Rev.} {\bf D 32}, 1248
(1985).
\item Note that the terms $\Phi^{\dagger} \Tilde{\Phi},~\Phi^{\dagger}
\Sigma,~\Phi^{\dagger}\Tilde{\Sigma}$ (where $\Tilde{\Phi}=\tau_2 \Phi^*
\tau_2$) are allowed after soft or
spontaneous  breaking of
the PQ symmetry.  These terms are sufficient to generate non--zero
phases for the vev's.
\item K.S. Babu and R.N. Mohapatra, in preparation.
\item J. Gasser and H. Leutwyler, {\it Phys. Rept.} {\bf 87}, 77 (1982).
\item K.S. Babu, {\it Z. Phys.} {\bf C 35}, 69 (1987);\\
K. Sasaki, {\it Z. Phys.} {\bf C 32}, 149 (1986);\\
B. Grzadkowski, M. Lindner and S. Theisen, {\rm Phys. Lett.} {\bf B198},
64 (1987).
\item N. Ushida et. al., {\it Phys. Rev. Lett.} {\bf 57}, 2897 (1986).
\item K.S. Babu and Q. Shafi, Bartol Preprint BA-92-27 (1992).
\item C. Jarlskog, {\it Phys. Rev. Lett.}, {\bf 55}, 1039 (1985).
\item J. Harvey, P. Ramond and D. Reiss, {\it Nucl. Phys.} {\bf B199},
223 (1982);\\
S. Dimopoulos, L. Hall and S. Raby, {\it Phys. Rev. Lett.} {\bf 68},
1984 (1992).\\
H. Arason et. al., Florida Preprint UFIFT-HEP-92-8;\\
G. Lazarides and Q. Shafi, {\it Nucl. Phys.} {\bf B350}, 179 (1991).
\end{enumerate}

\end{document}